\documentclass[useAMS,usenatbib]{mn2e}
\usepackage{natbib}
\usepackage{psfig}
\usepackage{color,graphicx} 
\usepackage{amsmath} 
\usepackage{amsfonts} 
\usepackage{amssymb} 
\usepackage{hyperref} 

\title[The SW~Sex enigma]{The SW~Sex enigma}
\author[V. S. Dhillon et al.]{V. S. Dhillon,$^{1}$\thanks{E-mail:
    vik.dhillon@sheffield.ac.uk} D. A. Smith,$^{2}$ T. R. Marsh$^{3}$ \\
$^{1}$Department of Physics and Astronomy, University of Sheffield, Sheffield S3 7RH, UK \\
$^{2}$Magdalen College School, Cowley Place, Oxford OX4 1DZ, UK \\
$^{3}$Department of Physics, University of Warwick, Coventry CV4 7AL, UK}

\begin{document}

\date{Submitted for publication on 2012 September 6. Revised on 2012
  October 24.}

\maketitle

\begin{abstract}
  The SW~Sex stars are a class of cataclysmic variables, originally
  identified because they shared a number of enigmatic properties --
  most notably, single-peaked emission lines instead of the
  double-peaked lines one would expect from their high-inclination
  accretion discs. We present high time-resolution spectrophotometry
  of the eclipsing nova-like variables SW~Sex and DW~UMa, two of the
  founding members of the SW~Sex class. Both systems show
  single-peaked Balmer and He\,{\small II} $\lambda$4686\AA\ emission
  lines that appear to originate from a region in the disc that lies
  close to, but downstream of, the bright spot. The emission-line
  light curves are consistent with the finding from X-ray and
  ultraviolet observations that we predominantly see the flared disc
  rim and the unobscured back portion of the disc in these systems. In
  DW~UMa, the He\,{\small II} $\lambda$4686\AA\ emission line
  originates from close to the white dwarf and exhibits flaring. Such
  flares have been used to argue for magnetically-channelled
  accretion, as in the intermediate polars, but the lack of a clear
  periodicity in the flares argues for a simpler model in which we are
  viewing the central regions of the disc through the non-uniform
  upper edge of a flared disc rim. We also observe narrow,
  blue-shifted, transient absorption features in the Balmer lines of
  DW~UMa, which we attribute to blobs of material ejected from the
  system, possibly by a magnetic propeller, that happen to be passing
  between us and the binary. Our results suggest that the solution to
  the SW~Sex enigma is a combination of dominant bright-spot emission
  and a self-occulting disc. We also propose a simplified
  classification scheme for nova-like variables.

\end{abstract}


\section{Introduction}

Cataclysmic variables (CVs) are close binaries consisting of a white
dwarf accreting material from a companion star via an accretion disc
or magnetic accretion stream. Nova-like variables (NLs) are those CVs
which have not been observed to undergo a nova or dwarf nova outburst.
The absence of dwarf nova outbursts in NLs is believed to be due to
their high mass transfer rates, producing ionised accretion discs in
which the disc instability mechanism is suppressed. For a review
of CVs and NLs, see \cite{warner95a} and \cite{hellier01}. 

Over two decades ago, a sub-class of NLs was identified that shared
common but inexplicable properties (e.g. \citealt{szkody90};
\citealt{dhillon90}; \citealt{thorstensen91a}).  These so-called
``SW~Sex stars'' exhibit deep continuum eclipses, implying a
high-inclination accretion disc. However, their emission lines show
single-peaked profiles and shallow eclipses, instead of the
deeply-eclipsing, double-peaked profiles one would expect
\citep{horne86}. Moreover, their emission lines do not share the
orbital motion of the white dwarf, and exhibit transient absorption
features, all indicative of a complex, possibly non-disc origin. One
could argue that determining the precise origin of the emission lines
is an unimportant detail, if it were not for the fact that the orbital
periods of SW~Sex stars lie tightly clustered just above the period
gap of CVs \citep{rodriguez-gil07b}. Hence, understanding the
processes responsible for making SW~Sex stars appear so different to
other classes of CV might also provide a clue as to the nature of the
period gap, which remains one of the most poorly understood links in
the chain of CV evolution \citep{knigge11}.

At the time of writing, there are 30 confirmed members of the SW~Sex
class\footnote{See D.~W.\ Hoard's Big List of SW~Sextantis Stars at
  \url{http://www.dwhoard.com/home/biglist} \citep{hoard03}.}.  The
membership criteria since the class was first recognised in 1991 has
changed somewhat (e.g. see \citealt{rodriguez-gil07a}).  Most notably,
large numbers of lower-inclination, non-eclipsing SW~Sex stars have
been found, making up approximately half of the population
\citep{rodriguez-gil07b}. Moreover, the SW~Sex stars are now known to
make up over half of the CV population {\em within} the 2--3 h period
gap, as well as dominating the population of CVs in the 3--4.5 h range
immediately above the gap \citep{rodriguez-gil07b}.

Several models\footnotemark[1] have been proposed to explain the
behaviour of SW~Sex stars, most of which include one or more of the
following ingredients: accretion disc wind, gas-stream overflow,
magnetic accretion, and obscuration by the disc rim. None are entirely
satisfactory -- see \cite{dhillon91}, \cite{hellier00}, \cite{hoard03}
and \cite{rodriguez-gil07b} for detailed discussions. With the aim of
discriminating between the various models, we obtained time-resolved,
spectrophotometric data of two of the founding members of the SW~Sex
class: DW~UMa (see \cite{hoard10} and references therein) and SW~Sex
(see \cite{dhillon97b}, hereafter DMJ97, and references therein). To
the best of our knowledge, the data presented in this paper are the
highest quality optical data published to date on these systems,
having been obtained with a telescope of twice the aperture of
previous studies.

\section{Observations and data reduction}
\label{sec:obsred}

On the nights of 1997 February 22 and 23 we obtained spectrophotometry
of DW~UMa and SW~Sex with the 4.2-m William Herschel Telescope (WHT)
on La Palma. The ISIS spectrometer \citep{carter93} with the R1200R
and R1200B gratings and the $1024\times1024$ TEK CCD chips gave a
wavelength coverage of approximately $4565-4970$\AA\ at $0.8$\AA\
(50\,km\,s$^{-1}$) resolution in the blue arm and $6420-6825$\AA\ at
$0.8$\AA\ (36\,km\,s$^{-1}$) resolution in the red arm. On the first
night we obtained 414 red and 420 blue spectra of DW~UMa, covering 3.0
orbits. On the second night we obtained 347 red and 345 blue spectra
of SW~Sex covering 2.3 orbits, and were also able to obtain a further
94 red and 94 blue spectra of DW~UMa.  The exposures were all 60\,s
with 15\,s dead-time for the archiving of data. The $1^{\prime\prime}$
slit was oriented to cover a nearby comparison star and wide slit exposures
were taken in order to correct for slit losses. Comparison arc spectra
were taken every $30-40$\,min to calibrate the spectrograph flexure.
Both nights were photometric, with the seeing ranging from
$1.5-2^{\prime\prime}$.

Data reduction was performed using TRM's {\sf pamela} and {\sf molly}
packages. We first corrected for pixel-to-pixel sensitivity variations
in the spatial and dispersion directions using sky and tungsten lamp
flat-fields, respectively. After sky subtraction, the data were
optimally extracted to give raw spectra of SW~Sex, DW~UMa and the
comparison stars. Arc spectra were then extracted from the same
locations on the detector as the targets. The wavelength scale for
each spectrum was interpolated from the wavelength scales of two
neighbouring arc spectra. The comparison stars were too faint to allow
for a wavelength-dependent slit-loss correction, so each SW~Sex and
DW~UMa spectrum was divided by the total flux in the corresponding
comparison star spectrum and multiplied by the total flux in the wide
slit comparison star spectrum. The observations were placed on an
absolute flux scale by using observations of the standard star
HD~19445 \citep{oke83} taken at the end of each night. The red spectra
contained significant telluric features which were removed using
spectra of the rapidly rotating B stars HR1037 and HR1786.

\section{Results}

\subsection{Average spectra}
\label{sec:avspec}

\begin{figure}
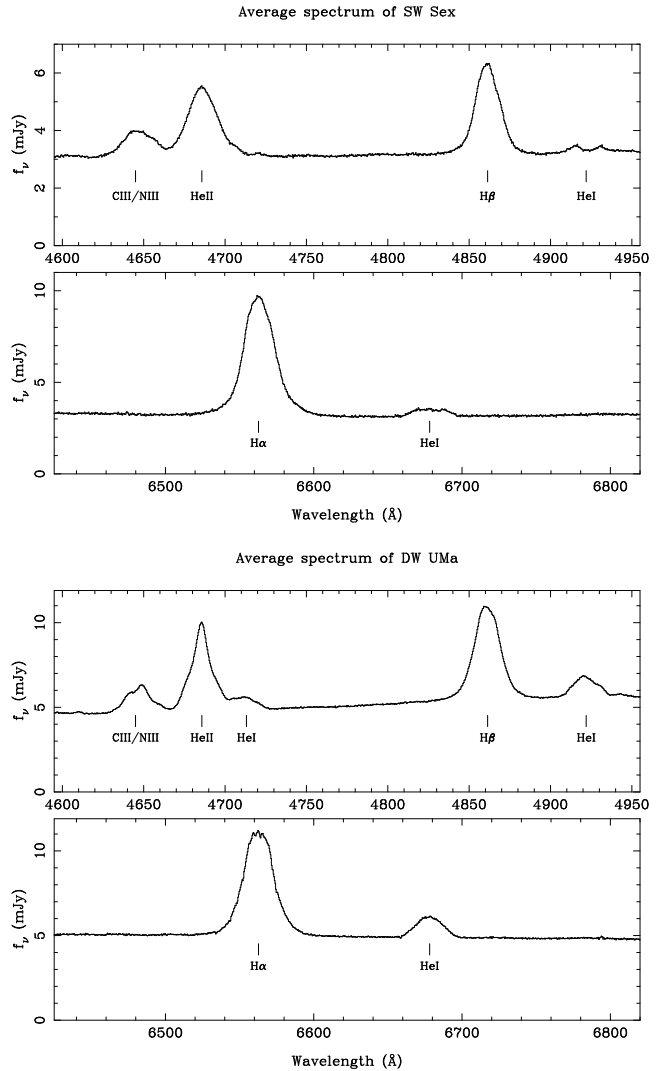

\centerline{\psfig{file=swav.ps,width=85mm,angle=-90}}
\ \newline \ 
\centerline{\psfig{file=dwav.ps,width=85mm,angle=-90}}
\caption{The average blue and red spectra of SW~Sex (top) and
DW~UMa (bottom), uncorrected for orbital motion.}
\label{fig:av}
\end{figure}

The averages of all blue and red spectra of SW~Sex and DW~UMa,
uncorrected for orbital motion, are displayed in Fig.~\ref{fig:av},
and in Table~\ref{tab:lines} we list fluxes, equivalent widths and
velocity widths of the most prominent lines measured from the average
spectrum.

SW~Sex shows broad emission lines due to the Balmer series,
He\,{\small I}, He\,{\small II} $\lambda$4686\AA\ and C\,{\small
  III}\,/\,N\,{\small III} $\lambda\lambda4640-4650$\AA. The average
spectrum in the blue is similar in appearance to that presented by DMJ97. The
continuum is weaker (around 3 mJy, compared to the 6 mJy reported
by DMJ97), but the emission lines are of similar strength. The
weak He\,{\small I} $\lambda$4922\AA\ line appears to be double
peaked, as expected for a line originating in a high-inclination
accretion disc, but the Balmer, He\,{\small II} and C\,{\small
  III}\,/\,N\,{\small III} lines are all single-peaked. There are no visible
secondary-star absorption features.  The spectrum of SW~Sex resembles
that of the other SW~Sex stars; a direct comparison of the spectrum of
SW~Sex with four other members of the class is given by
\cite{dhillon96}.

DW~UMa shows the same broad Balmer and He\,{\small I} lines as SW~Sex
(with the addition of the He\,{\small I} $\lambda$4713\AA\ line), as
well as the high-excitation lines of He\,{\small II} $\lambda$4686\AA\
and C\,{\small III}\,/\,N\,{\small III} $\lambda\lambda4640-4650$\AA.
The system appears to have come out of the low state observed by
\citeauthor{dhillon94} (\citeyear{dhillon94}; hereafter DJM94) and
returned to the high state observed by \cite{shafter88}: the continuum, at around
5\,mJy, and the return of the He\,{\small II} $\lambda$4686\AA\ line
to a strength comparable to that of H$\beta$ is in marked contrast to
the spectra shown by DJM94, in which there was no He\,{\small II}
$\lambda$4686\AA\ and the continuum was around 0.1\,mJy. H$\alpha$ and
C\,{\small III}\,/\,N\,{\small III} $\lambda\lambda4640-4650$\AA\ show
more structure in the average spectrum of DW~UMa than in SW~Sex, the
He\,{\small I} lines are not double-peaked, and the He\,{\small II}
$\lambda$4686\AA\ line also differs from that in SW~Sex, having a
narrow core on top of a wide base. No absorption lines from the
secondary star are detected.
 
\setlength{\tabcolsep}{2.5pt}
\begin{table}
\caption{Fluxes and widths of prominent lines in SW~Sex and DW~UMa, 
measured from the average spectrum.}
{\small
\begin{tabular}{lrrlc} 
& & & & \\
\multicolumn{1}{l}{Line} &
\multicolumn{1}{c}{Flux} & 
\multicolumn{1}{c}{EW} & 
\multicolumn{1}{c}{FWHM} & 
\multicolumn{1}{c}{FWZI} \\
\multicolumn{1}{l}{ } &
\multicolumn{1}{c}{$\times$ 10$^{-14}$} &
\multicolumn{1}{c}{\AA} & 
\multicolumn{1}{c}{km\,s$^{-1}$} & 
\multicolumn{1}{c}{km\,s$^{-1}$} \\
& erg\,cm$^{-2}$\,s$^{-1}$ & & & \\
 & & & & \\
SW~Sex & & & & \\
 & & & & \\
H$\alpha$ & $12.0\pm0.1$ & $54.5\pm0.2$ & $1050\pm50$  & $3800\pm200$ \\
H$\beta$  & $7.5\pm0.1$  & $20.0\pm0.1$ & $1050\pm50$  & $3700\pm200$ \\
He\,{\small II} $\lambda$4686\AA\ & $8.3\pm0.1$  & $18.5\pm0.5$ & $1350\pm50$  & \  --\\
C\,{\small III}/N\,{\small III} & $3.0\pm0.1$  & $7.0\pm0.5$	& $1500\pm50$  & \  --\\
He\,{\small I} $\lambda$4922\AA\  & $0.7\pm0.1$  & $2.0\pm0.1$	& $1350\pm100$ & $3200\pm200$ \\
He\,{\small I} $\lambda$6678\AA\  & $0.7\pm0.1$  & $3.2\pm0.1$ 	& $1200\pm50$  & $1700\pm200$ \\
 & & & & \\
DW~UMa & & & & \\
 & & & & \\
H$\alpha$ & $11.1\pm0.1$  & $31.9\pm0.1$ &  $1050\pm50$ & $3500\pm200$ \\
H$\beta$  & $14.3\pm0.1$  & $20.8\pm0.1$ &  $1100\pm50$ & $3700\pm200$ \\
He\,{\small II} $\lambda$4686\AA\ & $11.0\pm0.5$  & $16.5\pm0.5$ & \ \,$800\pm50$ &  \  --  \\
C\,{\small III}/N\,{\small III} & $3.9\pm0.1$   & $6.0\pm0.1$  &  $1050\pm50$ &  \  --  \\
He\,{\small I} $\lambda$4922\AA\ & $2.2\pm0.1$   & $3.2\pm0.1$  &
$1000\pm50$ & $1900\pm200 $ \\
He\,{\small I} $\lambda$6678\AA\ & $1.7\pm0.1$   & $5.2\pm0.1$  & \ \,$950\pm50$  & $1800\pm200$ \\
& & & & \\
\end{tabular}
}
\label{tab:lines}
\end{table}

\subsection{Light curves}
\label{sec:lc}

\begin{figure*}
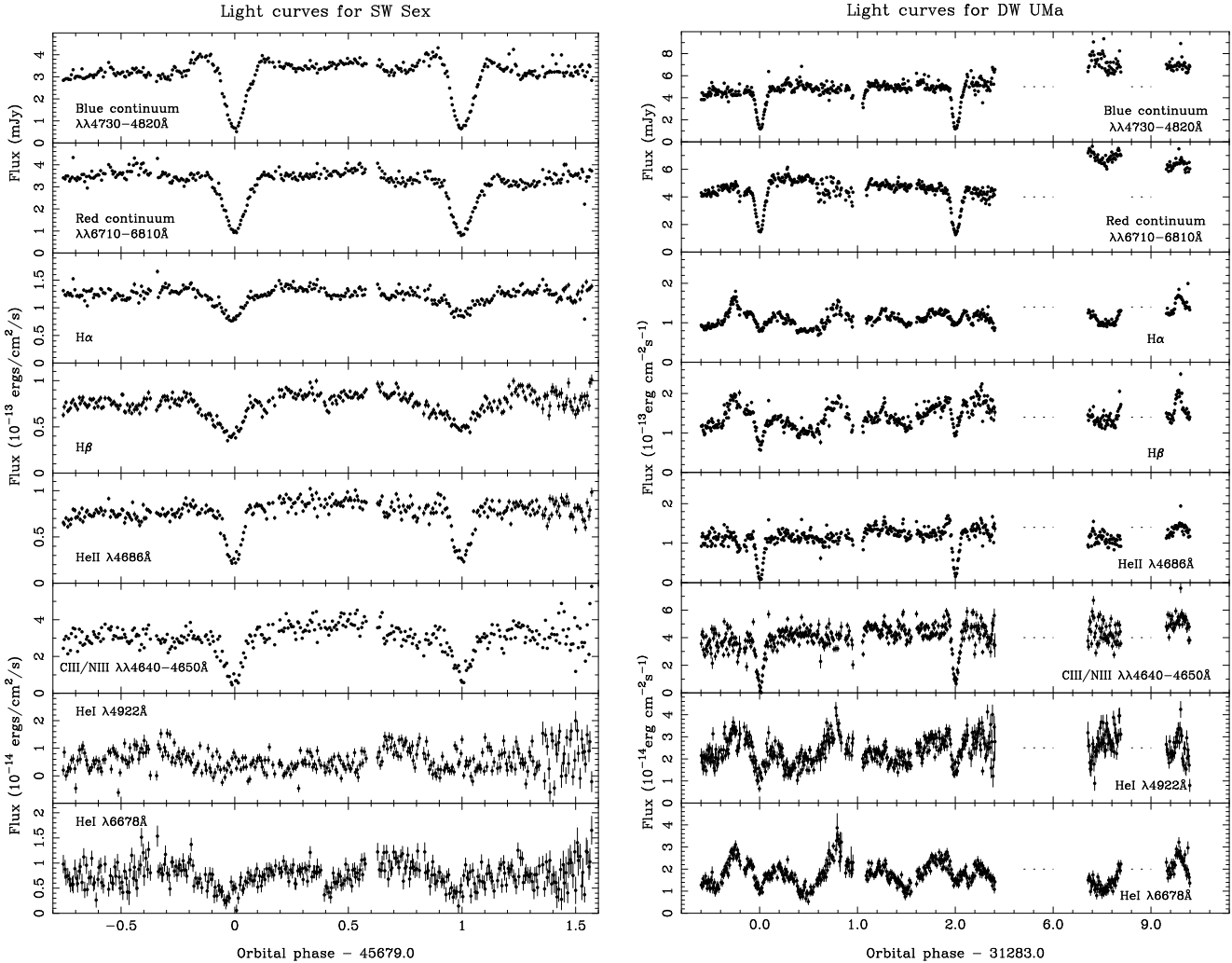

\centerline{\psfig{file=swlc.ps,width=85mm}\hspace*{0.5cm}\psfig{file=dwlc.ps,width=85mm}}
\caption{Continuum and emission-line light curves of SW~Sex (left) and DW~UMa (right).}
\label{fig:lc}
\end{figure*}

We recorded two eclipses each of SW~Sex (at $T_{\rm mid-ecl} = {\rm
  HJD}\, 2\,450\,503.5038\pm0.0004$ and $2\,450\,503.6388\pm0.0004$) and DW~UMa
(at $T_{\rm mid-ecl} = {\rm HJD}\, 2\,450\,502.4692\pm0.0004$ and
$2\,450\,502.7424\pm0.0004$). For SW~Sex, we combined these mid-eclipse timings
with those of \cite{penning84}, \cite{dhillon97b}, \cite{ashoka94} and
\cite{groot01} to obtain the following ephemeris:
\begin{equation}
\nonumber
\begin{array}{lrll}
T_{\rm mid-ecl} = & \hspace*{-0.2cm} {\rm HJD}\ 2\,444\,339.65057 
& \hspace*{-0.2cm}+\ 0.134938441 & \hspace*{-0.3cm} E \\
& \pm\ 0.00004 & \hspace*{-0.2cm} \pm\ 0.000000001. & \\
\end{array}
\end{equation}
For DW~UMa, we combined our eclipse timings with those of
\cite{shafter88} and \cite{dhillon94} to obtain the following
ephemeris:
\begin{equation}
\nonumber
\begin{array}{lrll}
T_{\rm mid-ecl} = & \hspace*{-0.2cm} {\rm HJD}\ 2\,446\,229.00696
& \hspace*{-0.2cm}+ \ 0.136606499 & \hspace*{-0.3cm} E \\
& \pm\ 0.00003 & \hspace*{-0.2cm} \pm\ 0.000000003. & \\
\end{array}
\end{equation}
The residuals from these fits show no evidence for period changes in
the decade-long baseline of eclipse timings. The above ephemerides
were used to place all of the data presented in this paper on
a phase scale. Note that a more up-to-date ephemeris for each target
has recently been presented by \cite{boyd12}.

Regions of the spectrum devoid of emission lines were selected in the
blue and red ($\lambda\lambda4730-4820$\AA\ and
$\lambda\lambda6710-6810$\AA, respectively). The blue and red
continuum light curves for SW~Sex and DW~UMa were then computed by
summing the flux in the above wavelength ranges. A third-order
polynomial fit to the continuum was subtracted from the blue and red
spectra and the emission-line light curves were then computed by
summing the residual flux between $\pm2000$\,km\,s$^{-1}$ for the
Balmer lines, $\pm1500$\,km\,s$^{-1}$ for the He\,{\small I} and
He\,{\small II} $\lambda$4686\AA\ lines and between
$\lambda\lambda4625-4665$\AA\ for C\,{\small III}\,/\,N\,{\small III}
$\lambda\lambda4640-4650$\AA. The resulting light curves for SW~Sex
and DW~UMa are plotted in Fig.~\ref{fig:lc}.

Our light curves of SW~Sex are qualitatively similar to those
presented by DMJ97 and \cite{groot01}.  The continuum shows a deep,
slightly asymmetrical eclipse, the ingress being steeper than the
egress, and the blue light curve having a deeper eclipse than the red,
consistent with the presence of a bright spot. There is the clear
signature of an orbital hump around phase $0.8-0.9$, as also seen by
\cite{rutten92b} and \cite{ashoka94}, caused by the changing aspect of
the bright spot. The only other feature in the continuum light curves
is the flickering seen in all CVs.  The eclipses of the Balmer lines
in SW~Sex are shallower than the continuum eclipses, and the asymmetry
is reversed, the egress being steeper than the ingress which begins
around phase 0.8. The eclipse appears to have two components, a narrow
central component from orbital phase $\sim -0.05$ to 0.05, which is
presumably the eclipse of the disc, and a broader, shallower component
from orbital phase $\sim -0.2$ to 0.2, which may be attributed to the
obscuration of the inner disc by a raised disc rim downstream of the
bright spot. The He\,{\small II} $\lambda$4686\AA\ and C\,{\small
  III}\,/\,N\,{\small III} $\lambda\lambda4640-4650$\AA\ eclipses are
quite deep and slightly narrower than the continuum light curves, but
show no other significant features. The light curves of both
He\,{\small I} lines show peculiar behaviour -- the $\lambda$4922\AA\
line is barely eclipsed and is strongest around orbital phase~0.7,
while the $\lambda$6678\AA\ line appears to be eclipsed early, with
mid-eclipse around orbital phase~0.95. 

There is no evidence of a reduction in flux around phase~0.5 in any of
the lines of SW~Sex, despite the fact that this is supposedly one of
the defining features of SW~Sex stars and SW~Sex itself is the
proto-type of the class \citep{thorstensen91a}. The system brightness
of SW~Sex during our observations was approximately the same as
observed by \cite{groot01}, who also failed to see the phase~0.5
absorption feature. This is in contrast to the observations of SW~Sex
presented by DMJ97, \cite{honeycutt86}, and \cite{szkody90}, during
which SW~Sex was a factor of two brighter and clearly showed the
phase~0.5 absorption feature. It is likely, therefore, that the
presence of phase~0.5 absorption is accretion-rate dependent.

The continuum light curves of DW~UMa show deep symmetrical eclipses
(the eclipse at orbital phase 31284.0 was missed because of cloud),
but there is no orbital hump apparent in these light curves, in
contrast to those of SW~Sex. The system appeared to brighten on the
second night, the continuum brightness out of eclipse rising from
around 5\,mJy during cycles $31282-31285$ to $7-8$\,mJy during cycles
31289 and 31292. The eclipses of the Balmer lines are shallow,
especially in H$\alpha$, in which the flux around phase~0.5 is lower
than during primary eclipse. The high-excitation lines of He\,{\small
  II} $\lambda$4686\AA\ and C\,{\small III}\,/\,N\,{\small III}
$\lambda\lambda4640-4650$\AA\ are deeply eclipsed, as in SW~Sex, and
again the eclipses of these lines are centred on phase zero and
narrower than the continuum eclipses, implying an origin coincident
with the inner disc region. The light curves of the He\,{\small I}
lines are similar to those of the Balmer lines. Note also that the
fluxes of all the lines do not change substantially on the second
night, despite the significantly increased continuum brightness.

\begin{figure*}
\centerline{\psfig{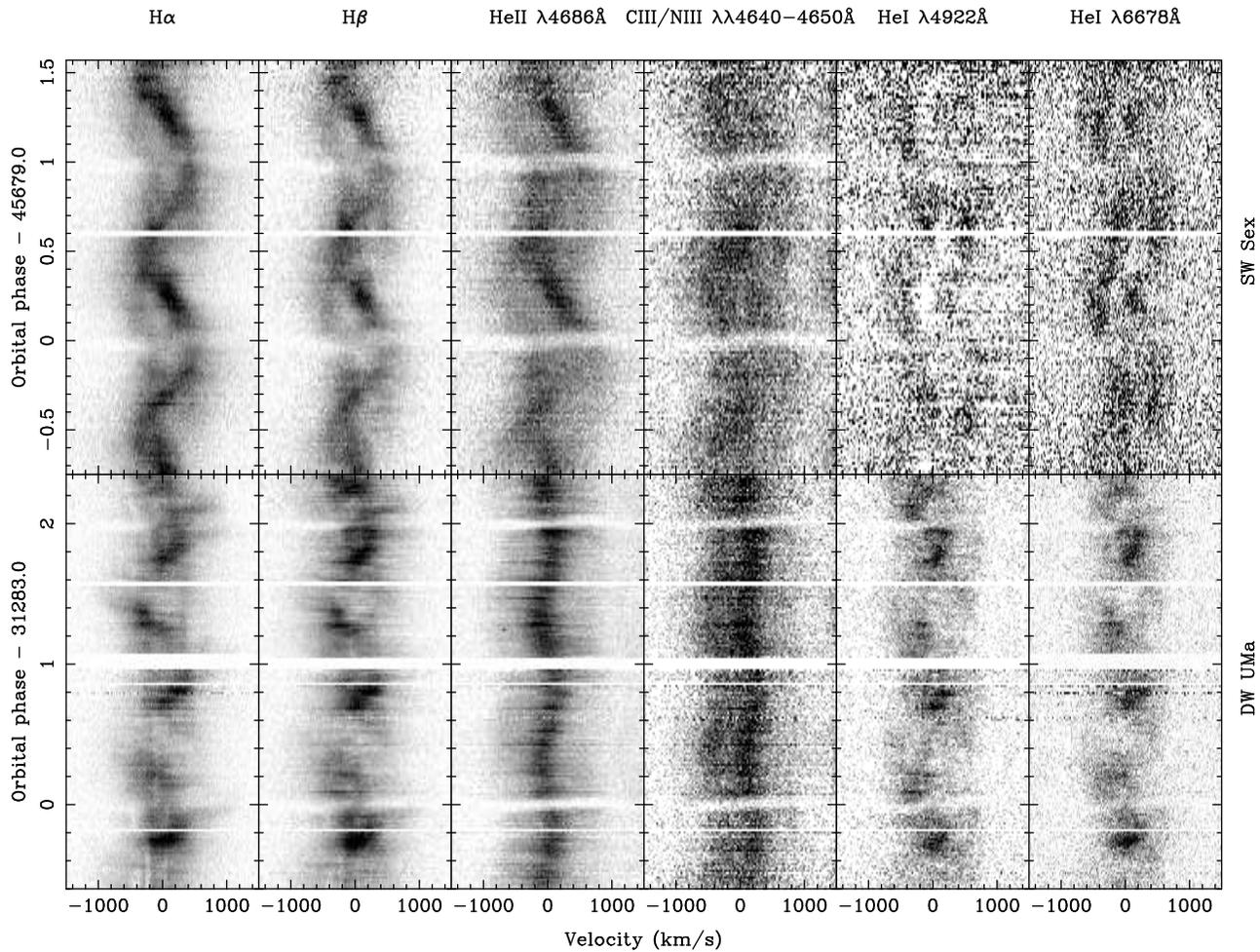}}
\caption{Trailed spectra of H$\alpha$, H$\beta$, He\,{\small II}
  $\lambda$4686\AA, C\,{\small III}\,/\,N\,{\small III}
  $\lambda\lambda4640-4650$\AA, He\,{\small I} $\lambda$4922\AA\ and
  He\,{\small I} $\lambda$6678\AA\ in SW~Sex (top) and DW~UMa
  (bottom).}
\label{fig:trail}
\end{figure*}

\subsection{Trailed spectra}
\label{sec:trail}

We subtracted the continuum from each of the spectra using a
third-order polynomial fit and then rebinned the spectra onto a
constant velocity-interval scale centred on the rest wavelengths of
the lines.  Fig.~\ref{fig:trail} shows the trailed spectra of the
H$\alpha$, H$\beta$, He\,{\small II} $\lambda$4686\AA, C\,{\small
  III}\,/\,N\,{\small III} $\lambda\lambda4640-4650$\AA, He\,{\small
  I} $\lambda$4922\AA\ and He\,{\small I} $\lambda$6678\AA\ lines in
SW~Sex (top) and DW~UMa (bottom).

In SW~Sex, the Balmer lines appear to have two major components. The
first and strongest starts at a velocity of $\sim300$\,km\,s$^{-1}$
just after eclipse and moves towards the blue. The second, weaker
component starts at a velocity of approximately $-400$\,km\,s$^{-1}$
and is in approximate anti-phase with the first component. The
dominant component in the He\,{\small II} $\lambda$4686\AA\ trail
shows a similar amplitude and phase to the strongest Balmer-line
component, but is clearly visible only through orbital phases
$0.1-0.5$ and shows a high velocity component $\sim90^\circ$ out of
phase with the expected motion of the white dwarf. The weaker lines
have less structure, though the double peaks of the He\,{\small I}
lines can be seen to follow an S-wave in approximate phase with the
expected motion of the white dwarf. All of the lines show evidence for
the eclipse of an accretion disc, with the blue wings of the lines
obscured before the red wings. The trailed spectra of SW~Sex show no
evidence for phase 0.5 absorption (see also Section~\ref{sec:lc});
apart from this, they are in good agreement with the trailed spectra
presented by DMJ97 and \cite{honeycutt86}, but of significantly higher
quality.

In DW~UMa, the Balmer lines are dominated by a single-peaked S-wave of
semi-amplitude $\sim 400$\,km\,s$^{-1}$, which is at its strongest
during $\phi\sim 0.1-0.4$ and $\phi\sim 0.7-0.9$. At other phases it
is almost entirely absent. There is a faint, high-velocity component
which shows up most clearly in the trail of the H$\alpha$ line,
running from around +1000\,km\,s$^{-1}$ at orbital phase~0.0 to
$-1000$\,km\,s$^{-1}$ at orbital phase~0.5. There are also
blue-shifted stationary absorption features running through the
Balmer-line trailed spectra, strongest in the first hour of
observation, but also faintly visible at some later times.  These are
discussed in more detail in Section~\ref{sec:tramlines}.  The
He\,{\small II} $\lambda$4686\AA\ line has a single S-wave component,
in phase with the expected motion of the white dwarf, with a lower
velocity amplitude than the Balmer lines
($\sim100$\,km\,s$^{-1}$). The blend of C\,{\small III}\,/\,N\,{\small
  III} $\lambda\lambda4640-4650$\AA\ shows a similar trail to that of
He\,{\small II} $\lambda$4686\AA.  Note also the quasi-periodic
flaring of the He\,{\small II} $\lambda$4686\AA\ line in DW~UMa,
visible as horizontal stripes in the trailed spectra. This is
discussed in more detail in Section~\ref{sec:flaring}.  Both
He\,{\small I} lines appear similar to the Balmer lines with strong
components at orbital phases $\phi\sim 0.1-0.4$ and $\phi\sim
0.7-0.9$. 

\begin{figure*}
  \centerline{\psfig{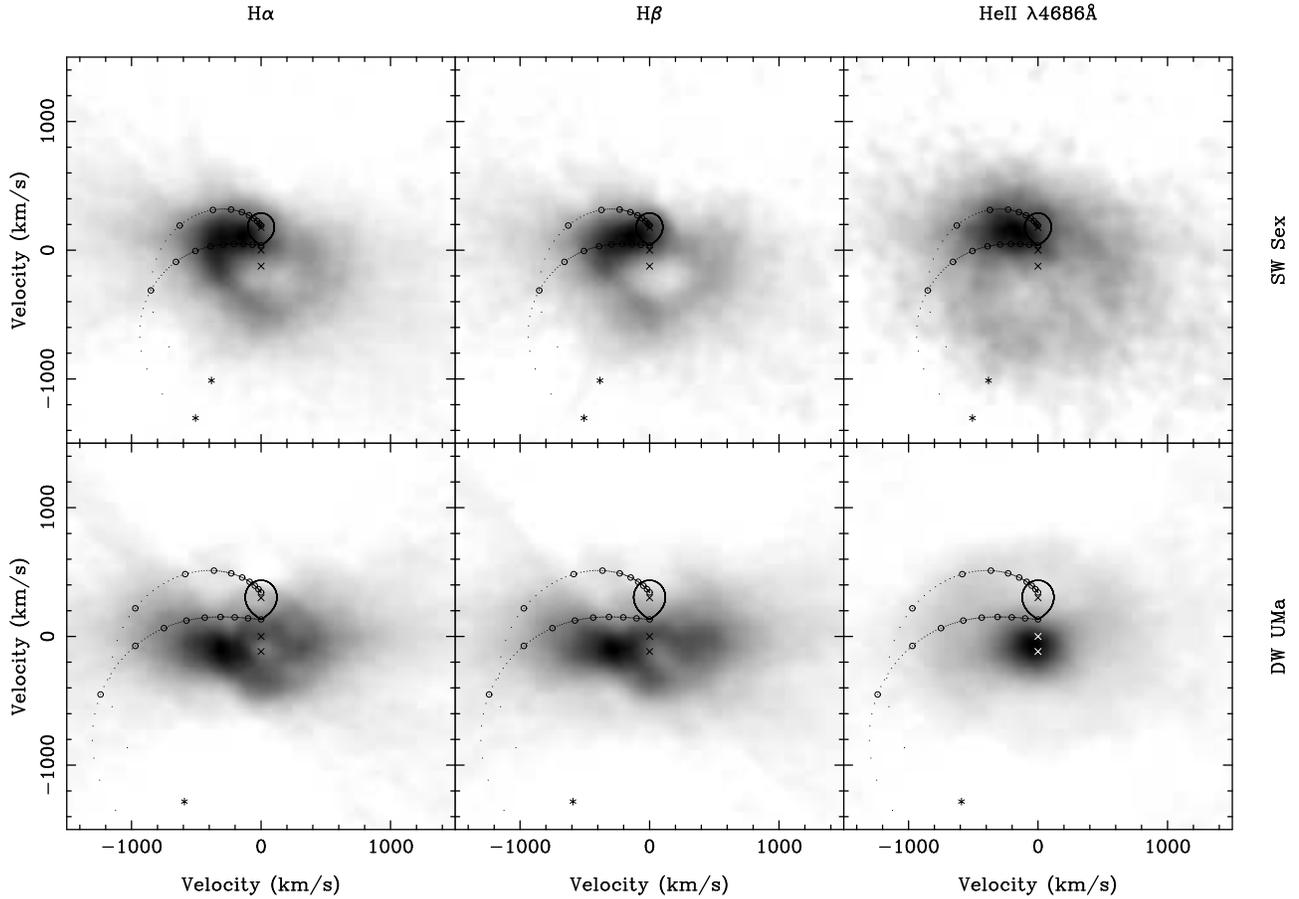}}
  \caption{Doppler maps of H$\alpha$, H$\beta$ and He\,{\small II}
    $\lambda$4686\AA\ in SW~Sex (top) and DW~UMa (bottom). The three
    crosses in each map are from top to bottom, the centres of mass of
    the secondary star, the system and the white dwarf. The predicted
    outline of the secondary star, the path of the gas stream (the
    lower curve) and the Keplerian velocity of the disc at the gas
    stream (upper curve) are marked, assuming $q=M_{2}/M_{1}=0.7$,
    $K_{W} + K_{R} = 300$\,km\,s$^{-1}$ for SW~Sex (representative
    values, following DMJ97 and {\protect\citealt{groot01}}) and $q=0.39$, $K_{W} +
    K_{R} = 418$\,km\,s$^{-1}$ for DW~UMa (calculated from the system
    parameters given by {\protect\citealt{araujo03}}). The series of circles
    along the path of the gas stream mark the distance from the white
    dwarf at intervals of $0.1R_{L1}$, where $1.0R_{L1}$ is the value
    at the secondary star.}
\label{fig:maps}
\end{figure*}

\subsection{Doppler maps}
\label{sec:maps}

The Doppler maps (see \cite{marsh01} for a review) of the three
strongest lines in SW~Sex and DW~UMa are shown in Fig.~\ref{fig:maps}.
These have been computed from the trailed spectra shown in
Fig.~\ref{fig:trail} with the eclipse spectra between orbital phases
$-0.1$ and $0.1$ removed.

In SW~Sex, the Balmer and He\,{\small II} $\lambda$4686\AA\ tomograms
are dominated by emission from a single component lying between the
gas stream velocity and the Keplerian velocity along the gas
stream. There is also some evidence for a ring of weak disc emission
with $v_{disc} \sim 300$\,km\,s$^{-1}$ in the Balmer-line tomograms,
but not in the He\,{\small II} $\lambda$4686\AA\ line. These tomograms
are quite different to those presented by DMJ97, which showed Balmer
emission from the inner face of the secondary star, and Balmer and
He\,{\small II} $\lambda$4686\AA\ emission from the bright spot.

In DW~UMa, the tomograms look very different. The Balmer lines show
hints of a low velocity ring-like structure -- the outer disc -- but
the position of peak intensity is around
$(-300,\,-100)$\,km\,s$^{-1}$, well away from the anticipated
ballistic stream trajectory. This emission in the $-v_x, -v_y$
quadrant is typical of SW~Sex stars, e.g. see \cite{kaitchuck94}. The
He\,{\small II} $\lambda$4686\AA\ line appears to be centred on or
near the white dwarf and shows no evidence for a ring of disc
emission.

\subsubsection{Tramlines in the trailed spectra}
\label{sec:tramlines}

\begin{figure}
\centerline{\psfig{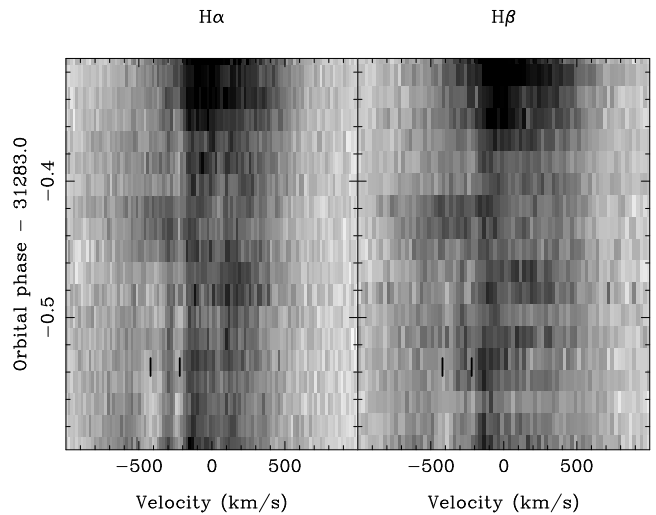}}
\caption{The tramlines at approximately $-420$ and
  $-220$\,km\,s$^{-1}$ in DW~UMa during the first hour of observation
  on the first night.}
\label{fig:tramlines}
\end{figure}

Looking carefully at the trailed spectra of H$\alpha$ in DW~UMa
(Fig.~\ref{fig:trail}), one notices vertical features running from
when observations commenced, at orbital phase 31282.4, until orbital
phase 31282.7, with possible reappearances at later times.  At first
we believed these narrow absorption lines at around --420 and
--220\,km\,s$^{-1}$ (and possibly also at +50\,km\,s$^{-1}$) were due
to telluric absorption, but they defied elimination. Careful
inspection of the H$\beta$ trailed spectrum, however, then revealed
similar features at similar velocities (at --420 and
--220\,km\,s$^{-1}$ but not at +50\,km\,s$^{-1}$). We are therefore
convinced that these blue-shifted `tramlines' are real features of the
spectra.  Fig.~\ref{fig:tramlines} represents our best enhancement of
these features. They do not appear in any of the other emission lines.
QU~Car has shown similar absorption features in H$\beta$, albeit at
higher velocities \citep{kafka12}. We discuss the origin of these
tramlines and their significance for models of SW~Sex stars in
Section~\ref{sec:discdw}.

\subsubsection{Flaring in the emission lines}
\label{sec:flaring}

\begin{figure}
\centerline{\psfig{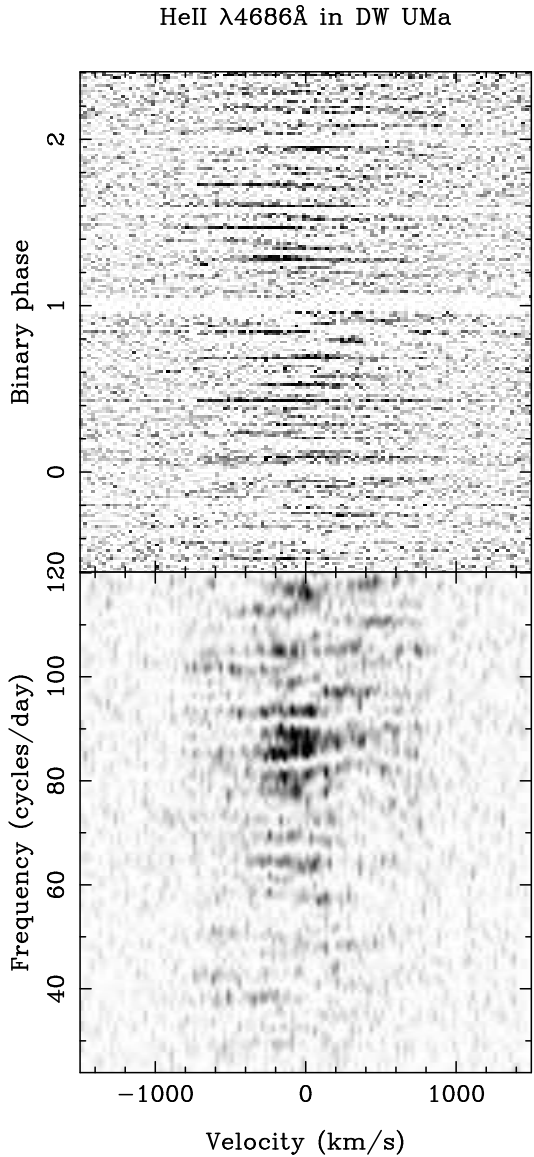}}
\caption{Top: Trailed spectrum of the He\,{\small II}
  $\lambda4686$\AA\ emission line in DW~UMa with the flares enhanced.
  Bottom: periodograms of the data in the upper panel as a function of
  wavelength, showing a quasi-periodicity around 88.5 cycles/day.  See
  section~\ref{sec:flaring} for details.}
\label{fig:flaring}
\end{figure}

The He\,{\small II} $\lambda$4686\AA\ trailed spectrum of DW~UMa
(Fig.~\ref{fig:trail}) shows quasi-periodic flaring similar to the features
seen more clearly in the trailed spectra of intermediate polars (IPs),
e.g. FO~Aqr \citep{marsh96} and DQ~Her \citep{bloemen10}. In the IPs,
the flaring is believed to be due to disc-reprocessing of X-ray beams
emitted from the magnetic poles of the asynchronously rotating white
dwarf. Emission-line flaring has already been observed in the SW~Sex
stars BT~Mon \citep{smith98}, LS~Peg \citep{rodriguez-gil01}, V533~Her
\citep{rodriguez-gil02}, RX\,J1643.7+3402 \citep{rodriguez-gil09},
BO~Cet and V380 Oph \citep{rodriguez-gil07a}, and interpreted as
evidence for magnetic accretion in these systems.

Because the flares are weak in DW~UMa, we attempted to enhance them by
removing the underlying line emission. This was achieved by median
filtering the trailed spectrum in the time direction using a filter of
significantly greater duration than the flares (15 spectra), and then
subtracting the result from the original trailed spectrum.  The
enhanced trailed spectrum is displayed in the top panel of
Fig.~\ref{fig:flaring}.

To determine if a periodicity is present, we removed the orbital
motion from the enhanced trailed spectrum plotted in the upper panel
of Fig.~\ref{fig:flaring} and then computed the Lomb-Scargle
periodogram at each wavelength \citep{press89}. The result is plotted
in the bottom panel of Fig.~\ref{fig:flaring}. The periodograms do
not show clear evidence for a single periodicity, but instead show a
broad set of peaks with a maximum around 88.5 cycles/day ($\sim
1000$\,s), indicating that the flaring is quasi-periodic, with
tentative evidence that the flaring is strongest in the blue side of
the He\,{\small II} $\lambda$4686\AA\ emission line. Unfortunately,
the flares are too weak to do a more sophisticated analysis than that
presented here.

\subsubsection{Radial velocities and system parameters}

We measured the radial velocities of the most prominent emission lines
in SW~Sex and DW~UMa using both the double-Gaussian method of
\cite{schneider80} and cross correlation with a single
Gaussian. Following the methods outlined by \cite{smith98}, we plotted
the resulting data on diagnostic diagrams \citep{shafter86} and
modified light-centre diagrams \citep{marsh88a} in order to account
for the observed phase shifts in the radial-velocity
curves. Unfortunately, we found that none of the lines could be
used to represent the motion of the white dwarf and hence it
is not possible to determine reliably the system parameters of SW~Sex
and DW~UMa with these data. DMJ97 and \cite{groot01} reached a similar
conclusion in their studies of SW~Sex and, to date, no reliable system parameters
exist for this object. The system parameters of DW~UMa, however, have
been constrained during low-state observations by DJM94 and
\cite{araujo03}, and via observations of the superhump period by
\cite{boyd09}.

\section{Discussion}
\label{sec:disc}

As the number of objects identified as SW~Sex stars has increased over
recent years, so the definition of SW Sex behaviour has become broader
(see \cite{rodriguez-gil07a} for a detailed description).  Also,
individual SW~Sex stars can show very different behaviour from one
observation to the next, due to the mass-accretion rate at the time of
observation, e.g. compare the low-state observations of DW~UMa (DJM94)
with those of the high state (this paper). Even the proto-typical NL
UX~UMa has now been shown to exhibit SW~Sex-like behaviour
\citep{neustroev11}, leading one to suspect that all NLs could be
classified as SW~Sex stars if one looks long and hard enough. We are
faced with having to explain a wide range of variable, high-$\dot{M}$
related phenomena, making it difficult to produce a single, simple,
self-consistent model for SW Sex stars.

Before discussing each object in detail, it is important to clarify
the NL classification scheme. In the scheme described by
\cite{warner95a}, the non-magnetic NLs are sub-divided into two main
classes, UX~UMa stars and RW~Tri stars, where the former have
persistent, broad Balmer absorption-line spectra and the latter have
pure emission-line spectra. As \cite{warner95a} notes, this difference
is probably a matter of inclination (see also Fig.~4 of
\citealt{dhillon96}). However, it is probably also something to do
with $\dot{M}$ at the time of observation, e.g. UX~UMa itself has been
shown to exhibit an emission-line spectrum at certain times
\citep{neustroev11}. Hence the sub-classification into UX~UMa and
RW~Tri stars is not useful. Instead, we propose that all non-magnetic
NLs (excluding the double-degenerate AM~CVn systems) should be called
UX~UMa stars. Some UX~UMa stars show low states, in which case they
should also be referred to as VY~Scl stars (or anti-dwarf novae), and
some UX~UMa stars show SW~Sex characteristics, in which case they
should also be referred to as SW Sex stars. Hence a particular UX~UMa
star can be a VY~Scl star {\em and} an SW~Sex star -- DW~UMa is one
such example.

\subsection{SW Sex}
\label{sec:discsw}

The observations of SW~Sex presented in this paper appear to be
relatively straight-forward to interpret. The Balmer and He\,{\small
  II} $\lambda$4686\AA\ emission lines are single peaked because they
are formed in a single region in the disc, close to where the material
in the gas stream and disc merge. The resulting radial-velocity curves
reflect the motion of this region about the centre of mass, and hence
exhibit phase shifts with respect to photometric minimum. The phase
0.5 absorption is not present in our observations of SW~Sex, most
probably due to a lower mass accretion rate (see
Section~\ref{sec:lc}).

The only difficulty is in explaining the emission-line light curves of
Fig.~\ref{fig:lc} -- if line emission from the vicinity of the bright
spot is dominant, one would expect a deep, narrow eclipse that is not
centred on phase 0. This is not observed in Fig.~\ref{fig:lc}, but
there are two caveats. First, the bright spot is not necessarily as
dominant as it appears -- sharp features tend to dominate Doppler maps
making it difficult to assess their overall significance. Second, the
emission-line light curves presented in Fig.~\ref{fig:lc} show the sum
of the line flux at each phase, whereas it is more instructive to
inspect how the fluxes of the different components that make up each
emission line vary with phase in the trailed spectra
(Fig.~\ref{fig:trail}).  The latter figure shows that the eclipse
appears to be dominated by two components: the eclipse of an accretion
disc centred on phase zero where, as expected, the blue-shifted region
of the disc is eclipsed prior to the red-shifted region, and the
eclipse of the dominant single-peaked bright-spot component, which
starts earlier and finishes later than the accretion disc eclipse. The
eclipse of the bright-spot component is too wide to be explained by
obscuration by the secondary star. Instead, we propose that the bright
spot overflows the disc edge and is obscured around primary eclipse by
a raised disc rim that only allows the back portion of the disc to be
seen. Such a self-occulting disc has already been proposed for DW~UMa
\citep{knigge00} and the SW~Sex star V348~Pup \cite{froning03}, and
would explain why the disc appears to show only a weak, narrow
double-peaked profile. The relatively shallow depth of the Balmer-line
eclipses can then be understood if the bright-spot emission region is
vertically extended, as proposed by \cite{groot01} and \cite{hoard03},
and modelled by \cite{kunze01}, or if there is a region of line
absorption in the disc that is obscured by the secondary star, as
proposed by DMJ97 and observed by \cite{groot01}.

\subsection{DW UMa}
\label{sec:discdw}

The observations of DW~UMa presented in this paper are quite different
to those of SW~Sex; in fact, DW~UMa looks more like an SW~Sex star
than SW~Sex itself, with Balmer emission from the $-v_x, -v_y$
quadrant in the Doppler maps and Balmer light curves showing only
shallow eclipses and evidence for phase 0.5 absorption. Our
observations of DW~UMa have also presented us with two new features
that have not been observed before in this object: flaring and
tramlines in the trailed spectra. We must also take into account the
fact that, unlike SW~Sex, DW~UMa has been extensively studied at
ultraviolet and X-ray wavelengths (\citealt{knigge00},
\citealt{araujo03}, \citealt{knigge04}, \citealt{hoard03},
\citealt{hoard10}) -- these observations have shown conclusively that
the inner disc of DW~UMa is obscured by the flared edge of the disc.

\begin{figure}
\centerline{\psfig{file=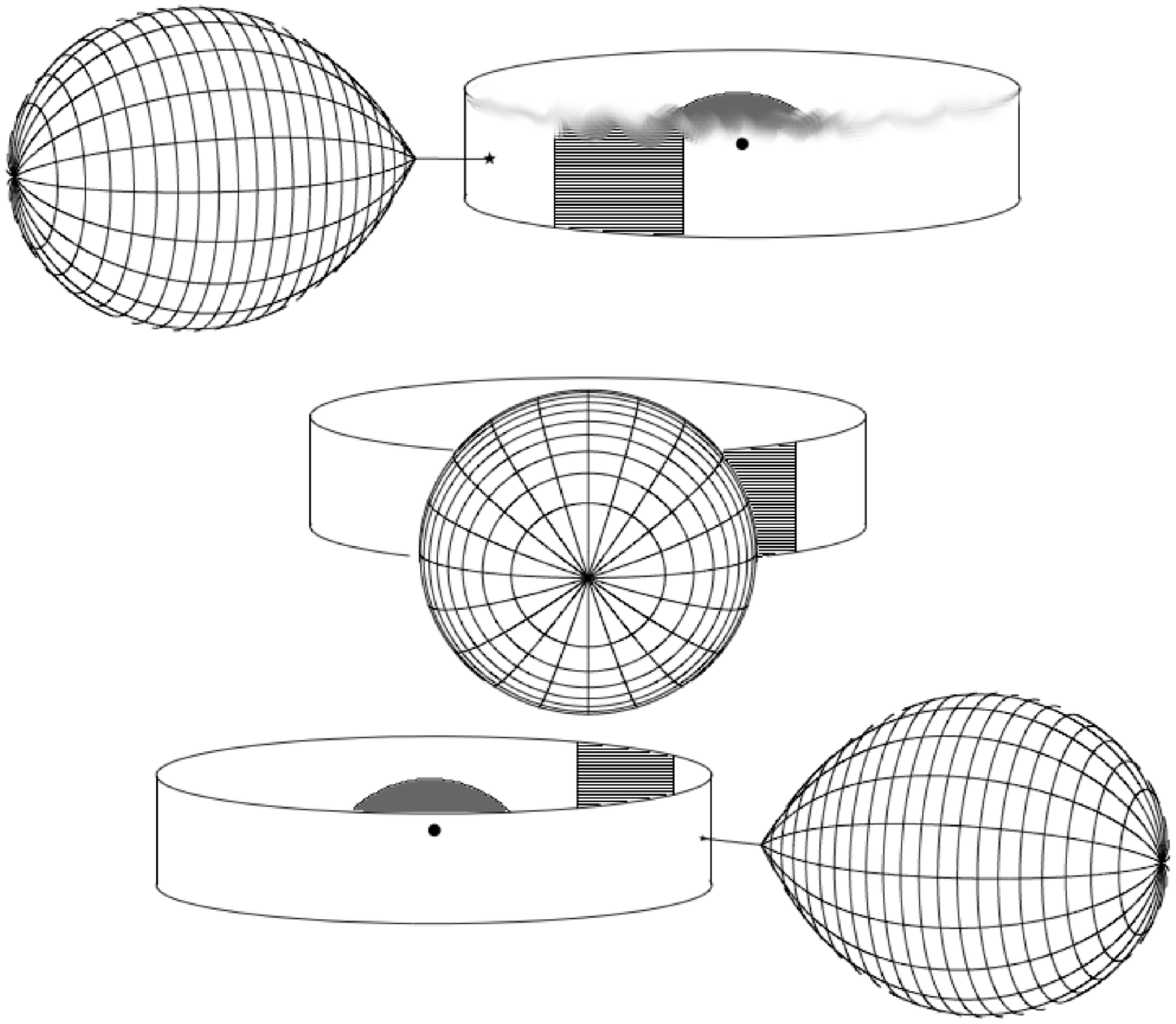,width=110mm,angle=270}}
\caption{Schematic showing the model of DW UMa proposed in
  section~\ref{sec:discdw}, at binary phases 0.8 (top), 1.0 (middle)
  and 1.2 (bottom). The plot is to scale, using the system parameters
  of {\protect\cite{araujo03}} and DJM94 ($q=0.39$, $i=82^{\circ}$,
  $R_{D}=0.8 L_{1}$), and assuming the ratio of the disc height to
  disc radius is 0.2. Note that the white dwarf is obscured by the upper
  edge of the accretion disc and is not visible -- it has, however,
  been included in the plot as a small black dot (not to scale) for
  reference. The key components of our model are the dominant Balmer
  emission region downstream of the bright spot on the disc rim
  (shaded box), the He\,{\small II} $\lambda$4686\AA\ emission region
  close to the white dwarf (shaded arc), and the non-uniform disc
  edge. For clarity, the non-uniform disc edge is only indicated on
  the near-side upper disc edge at phase 0.8, but actually extends all
  the way round the disc edge, on both the upper and lower edges, and
  is visible at all orbital phases. Note how the Balmer emission
  region on the disc edge is visible when viewed from both the front
  (at phase 0.8) and the back (at phase 1.2). Note also how the
  He\,{\small II} $\lambda$4686\AA\ region is viewed through the
  non-uniform disc edge (giving rise to the flaring), and how it is
  visible above the disc rim at all orbital phases except around
  primary eclipse, when it is obscured by the secondary star.}
\label{fig:dw_schematic}
\end{figure}

The above observational constraints lead us to the following model for
DW~UMa, shown pictorially in Fig.~\ref{fig:dw_schematic}.  The Doppler
maps suggest that the bulk of the Balmer emission originates in a
single post-shock region that lies downstream (along the disc edge)
from the bright spot. Theoretical estimates by \cite{spruit98} suggest
that it is feasible for this region to lie many tens of degrees in
disc azimuth from the impact zone. Since our view of the disc is
dominated by the rim, we see only very weak, narrow, double-peaked
underlying emission from the face of the disc, as modelled by
\cite{knigge00}. Hence this model explains the single-peaked lines and
the phase-shifted radial-velocity curves (or equivalently, their
location in the $-v_x, -v_y$ quadrant of the Doppler maps). To explain
the Balmer-line light curves, we again urge the reader to inspect the
variation in the strength of the dominant single-peaked component in
the trailed spectra (Fig.~\ref{fig:trail}) rather than the summed line
flux shown in Fig.~\ref{fig:lc}. It can be seen that, as expected in
our model, the line flux peaks around phase 0.8 when the post-impact
region passes across the line of sight, and then gradually diminishes
in strength due to foreshortening and obscuration by the secondary
star. There is not a deep eclipse of the Balmer emission because it is
dominated by emission from the disc rim and a substantial fraction of
the rim remains visible at phase 0. The line flux then grows in
strength again as we begin to see the back side of the post-impact
region around phase 0.2, and peaks around phase 0.3 as the region
passes across the line of sight on the back side of the disc. One
would then expect a gradual reduction in line flux around phase 0.5
due to foreshortening, before it comes to a maximum again around phase
0.8. This is not observed. Instead, we see a dramatic drop in line
flux around phase 0.5, which we attribute to absorption by material in
our line of sight to the post-impact region -- see below.

One of the major differences between SW~Sex and DW~UMa is the location
of the He\,{\small II} $\lambda$4686\AA\ emission. In SW~Sex it is
more or less coincident with the Balmer emission in the outer disc,
whereas in DW~UMa the Doppler map and deep, narrow eclipse indicate
that the He\,{\small II} $\lambda$4686\AA\ emission originates from
close to the white dwarf. We know that we don't see the white dwarf in
the high state of DW~UMa due to obscuration by the flared disc rim
\citep{knigge04}, so the He\,{\small II} $\lambda$4686\AA\ most likely
originates from a region lying above the inner disc, possibly in an
accretion disc wind (as modelled by \citealt{hoare94}). This region
must be low enough to be almost totally eclipsed by the secondary star
at phase 0, yet high enough to be visible above the disc rim at all
other phases, which means that we must be viewing the He\,{\small II}
$\lambda$4686\AA\ emission region through the upper edge of the disc
rim. This edge is unlikely to be uniform, and hence at some phases we
will see more of the emission region than at other phases, resulting
in the apparent flaring that we have observed in the line. Of course,
obscuration by a non-uniform disc edge would not explain the flaring
that has apparently been seen in lower-inclination SW~Sex stars (see
Section~\ref{sec:flaring}).

An alternative explanation for the origin of the He\,{\small II}
$\lambda$4686\AA\ emission in DW~UMa is in a magnetic accretion
curtain above the plane of the disc but close to the white dwarf
(\citealt{williams89};\citealt{dhillon91}; \citealt{hoard03}).  This
model would explain why the line is single peaked and deeply eclipsed,
and also provides a possible explanation for the flaring: an accretion
hot spot on an asynchronously rotating white dwarf that periodically
illuminates the He\,{\small II} $\lambda$4686\AA\ emission region in
the curtain.  The big problem with this interpretation is the lack of
coherence in the pulsations.

There remain two observed features in DW~UMa that we have not yet
addressed: the phase 0.5 absorption and the tramlines. Both are
absorption features and both are only visible in the Balmer lines. The
tramlines are transient, narrow, blue-shifted features, whereas the
phase 0.5 absorption repeats every orbit and occurs over a much wider
range of velocity. The transient nature of the tramlines rules out
absorption by widespread circumbinary gas, such as may have been
ejected during ancient nova outbursts. Instead, the tramlines are more
likely to be due to absorption by blobs of material recently ejected
from the system which happen to be passing between us and the CV and
are moving outwards. \cite{honeycutt86} invoked disc-wind material
overflowing the $L_3$ point to explain the phase 0.5 absorption in
SW~Sex stars, and perhaps the same mechanism accounts for the
tramlines. However, in this model, one would expect any absorption to
occur after phase 0.5, whereas both the tramlines and phase 0.5
absorption start around phase 0.4. Another mechanism to produce the
absorbing blobs of material is a magnetic propeller \citep{wynn97},
where accreting material encounters the rapidly spinning magnetosphere
of the white dwarf and is ejected from the binary. In fact,
\cite{horne99} has proposed that a magnetic propeller can also explain
the phase 0.5 absorption, where we view the system through the exit
stream. Alternatively, the phase 0.5 absorption might be due to
obscuration by a vertically raised structure in the disc, such as an
overflowing gas stream -- see \cite{hellier00} and \cite{hoard98}.

\section{Conclusions}
\label{sec:conc}

Using high-quality optical spectrophotometry, we have presented a
detailed comparison of two of the founding members of the SW~Sex class
of CVs: SW~Sex itself and DW~UMa. We find that our view of both
systems is dominated by the disc rim, which is largely responsible for
the variations in line flux that we observe. The Balmer emission
appears to originate from a region around the bright spot: in DW~UMa
this region lies further downstream from the impact point, and closer
to the disc edge, than in SW~Sex, indicative of differing mass
transfer rates and/or disc densities in these two systems (see
\cite{hoard98} for a discussion).  We have discovered flaring in the
He\,{\small II} $\lambda$4686\AA\ emission line and transient,
blue-shifted tramlines in the Balmer lines of DW~UMa. Both features
can be explained if the system is magnetic, but the lack of a clear
periodicity in the flares leads us to prefer a simpler model in which
we are viewing the central regions of the disc through the non-uniform
upper edge of the flared disc rim. The most marked difference between
the two systems is the location of the He\,{\small II}
$\lambda$4686\AA\ emission. In SW~Sex, this line seems to originate
from the bright spot region, whereas in DW~UMa it originates from
close to the white dwarf. Again, this could be indicative of a higher
mass-transfer rate in SW~Sex.

\section*{Acknowledgements}

We thank Paul Groot for communicating his eclipse timings, and Keith
Horne and Stuart Littlefair for useful discussions. We would also like
to thank the anonymous referee for comments that improved the paper.
The WHT is operated on the island of La Palma by the Isaac Newton
Group in the Spanish Observatorio del Roque de los Muchachos of the
Instituto de Astrofisica de Canarias.

\bibliographystyle{mn2e}
\bibliography{abbrev,refs}

\end{document}